\documentstyle[12pt,aasms4]{article}

\def \etal     {et al.}
\def \eg       {e.\,g.}
\def \vLSR     {\hbox{${v_{\rm LSR}}$}}
\def \delv     {\hbox{$\Delta v_{1/2}$}}
\def \TMB      {\hbox{$T_{\rm MB}$}}
\def \Tsys     {\hbox{$T_{\rm sys}$}}
\def \kms      {\hbox{${\rm km\,s}^{-1}$}}                
\def \Kkms     {\hbox{${\rm K\,km\,s}^{-1}$}}             
\def \AMM      {\hbox{NH$_3$}}                            
\def \HthCN    {\hbox{H$^{13}$CN}}                        
\def \HCfiN    {\hbox{HC$^{15}$N}}                        
\def \C#1      {\hbox{$^{#1}$C}}                          
\def \N#1      {\hbox{$^{#1}$N}}                          
\def \O#1      {\hbox{$^{#1}$O}}                          
\def \ISOC     {\hbox{$^{12}$C$/^{13}$C}}                 
\def \ISON     {\hbox{$^{14}$N$/^{15}$N}}                 
\def \ISOB     {\hbox{$^{18}$O$/^{17}$O}}                 
\def \bra#1    {$\left\{\makebox{\rule[-#1ex]{0pt}{#1ex}}\right.$}
\def \ket#1    {$\left.\makebox{\rule[-#1ex]{0pt}{#1ex}}\right\}$}
\def \uspace#1 {\makebox{\rule[#1ex]{0pt}{2ex}}}
\def \dspace   {\makebox{\rule[-2ex]{0pt}{2ex}}}

\begin{document}

\title{The detection of extragalactic $^{15}$N: Consequences for
       nitrogen nucleosynthesis and chemical evolution\altaffilmark{1,2}}

\author{Yi-nan Chin}
\affil{Institute of Astronomy and Astrophysics, Academia Sinica,
       P.O.Box 1-87, Nankang, 11529 Taipei, Taiwan;
       einmann@asiaa.sinica.edu.tw}

\author{Christian Henkel}
\affil{Max-Planck-Institut f\"ur Radioastronomie,
       Auf dem H\"ugel 69, D-53121 Bonn, Germany;
       p220hen@mpifr-bonn.mpg.de}
\affil{European Southern Observatory, Casilla 19001, Santiago 19, Chile}

\author{Norbert Langer}
\affil{Institut f\"ur Physik, Universit\"at Potsdam,
       Am Neuen Palais 10, D-14469 Potsdam, Germany;
       ntl@astro.physik.uni-potsdam.de}

\and

\author{Rainer Mauersberger}
\affil{Steward Observatory, The University of Arizona,
       Tucson, AZ 85721, U.S.A.;
       mauers@as.arizona.edu}

\altaffiltext{1}{Based on observations with the Swedish-ESO
                 Submillimeter Telescope (SEST) at the European Southern
                 Observatory (ESO), La Silla, Chile}
\altaffiltext{2}{Please send offprint request to
                 Y.-N. Chin, ASIAA, Taiwan; einmann@asiaa.sinica.edu.tw}

\begin{abstract}

   Detections of extragalactic \N15 \ are reported from observations of the rare
   hydrogen cyanide isotope \HCfiN\ toward the Large Magellanic Cloud (LMC) and
   the core of the (post-) starburst galaxy NGC\,4945.
   Accounting for optical depth effects, the LMC data from
   the massive star-forming region N113 infer a \ISON\ ratio of 111 $\pm$ 17,
   about twice the \ISOC\ value.
   For the LMC star-forming region N159HW and for the
   central region of NGC\,4945, \ISON\ ratios are also $\approx$ 100.
   The \ISON\ ratios are smaller than all interstellar nitrogen isotope ratios
   measured in the disk and center of the Milky Way, strongly supporting
   the idea that \N15 \ is predominantly synthesized by massive stars.
   Although this appears to be in contradiction with standard stellar evolution
   and nucleosynthesis calculations, it supports recent findings of abundant
   \N15 \ production due to rotationally induced mixing of
   protons into the helium-burning shells of massive stars.

\end{abstract}


\newpage
\section{Introduction}
 \label{sec:15N-Introduction}

   Carbon, nitrogen, and oxygen, the `CNO-elements',
   are among the most abundant species after hydrogen and helium.
   Being formed by p- and $\alpha$-capture reactions in the interior of stars,
   the CNO nuclei are partially released by means of stellar winds, planetary
   nebula ejecta, and supernova explosions into the interstellar medium (ISM).
   $^{12}$C and \O16 \ and, qualitatively, \C13 \ and \O17 \
   nucleosynthesis appears to be understood (\eg, Wilson \& Matteucci 1992;
   Henkel \& Mauersberger 1993; Henkel \etal\ 1994b; Prantzos \etal\ 1996).
   The dominant cooking site of the rare nitrogen isotope,
   \N15 , is however not yet known.
   While \N14 \ is produced in both high and lower mass stars,
   it is not yet clear whether \N14 \ or \N15 \ is the more `primary' isotope
   (see Sect.\,\ref{sec:15N-Discussion}).

   To obtain new constraints for nitrogen nucleosynthesis, we measured
   \ISON\ abundance ratios in so far unexplored molecular environments.
   The molecular studies by Johansson \etal\ (1994),
   Chin \etal\ (1996, 1997, 1998), and Heikkil\"a \etal\ (1997, 1998)
   demonstrate that it is possible to detect rare molecular species
   in the Magellanic Clouds.
   Since such metal poor environments may also characterize cosmologically
   relevant sources at high redshift, we searched for \HCfiN\ in three
   Magellanic star-forming regions showing prominent
   H$^{12}$C$^{14}$N (hereafter HCN) emission.
   In view of the low \ISON\ ratios predicted by Henkel \& Mauersberger (1993)
   and Henkel \etal\ (1994b) for starbursts, we also searched for
   \HCfiN\ in the southern (post-) starburst galaxy NGC\,4945.

\section{Observations}
 \label{sec:15N-Observations}

   The data were taken in September and November 1997 and in January and July
   1998 using the 15-m Swedish-ESO Submillimetre Telescope (SEST)
   at La Silla, Chile.
   A 3\,mm SIS receiver yielded overall system temperatures,
   including sky noise, of order \Tsys\ = 250\,K on
   a main beam brightness temperature (\TMB) scale.
   The backend was an acousto-optical spectrometer (AOS).
   A channel separation of 42\,kHz corresponding to 0.14\,\kms\
   at 88\,GHz was employed for the observations toward the Magellanic Clouds,
   while the low resolution spectrometer with a channel separation
   of 0.7\,MHz (or 2.4\,\kms) was used for NGC\,4945.
   The antenna beamwidth was 55\arcsec\ at the observed line
   frequencies taken from Lovas (1992).

   The observations were carried out in a dual beam-switching mode (switching
   frequency 6\,Hz) with a beam throw of 11\arcmin 40\arcsec\ in azimuth.
   All spectral intensities were converted to a \TMB\ scale,
   correcting for a main beam efficiency of 0.76
   (L.B.G.~Knee, L.-\AA.~Nyman, A.R.~Tieftrunk, priv.~comm.).
   Calibration was checked by monitoring on Orion KL and NGC\,4945 and was found
   to be consistent between different observation periods within $\pm$\,10\%.
   The pointing accuracy, obtained from measurements of the SiO masers
   R\,Dor and W\,Hya, was better than 10\arcsec.

\section{Results}
 \label{sec:15N-Results}

   Toward the prominent LMC star-forming regions N113 and N159HW
   we have detected the $J=1-0$ emission lines of hydrogen cyanide (HCN) and
   its rare isotopic species \HthCN\ and \HCfiN.
   Although they were also detected toward the core of the (post-) starburst
   galaxy NGC\,4945 (\eg, Koornneef 1993), only HCN was seen in the
   SMC star-forming region LIRS\,36 (see Chin \etal\ 1998).
   Spectra and line parameters obtained from Gaussian fits are displayed in
   Fig.\,\ref{fig:15N-spectra} and Table\,\ref{tbl:15N-parameter}.
   Observed line intensity ratios, derived opacities, and isotopic abundance
   ratios (\ISOC\ and \ISON) are given in Table\,\ref{tbl:15N-ratio}.

   Apart from optical depth effects, integrated line intensity ratios should be
   identical to isotope ratios within the observational errors,
   since rotational constants of the various HCN isotopomers are similar.
   A fractionation of nitrogen isotopes via charge exchange reactions, as
   expected in the case of carbon (see Watson \etal\ 1976), must be negligible,
   because the nitrogen ionization potential is higher than that of hydrogen;
   N$^{+}$ abundances are small in molecular clouds.
   For the two LMC star-forming regions and
   for NGC\,4945 we then find \ISON\ $\approx$ 100.
   In the case of N113, the \ISON\ ratio does not
   depend on an assumption on \ISOC.
   Instead, \ISON\ is {\em directly} derived from the
   HCN/\HCfiN\ integrated line temperature ratio and from a
   fit to the HCN hyperfine components (see Table\,\ref{tbl:15N-ratio}).
   Toward N159HW, the HCN main line also appears to be almost optically
   thin but a larger intrinsic linewidth makes the fit less convincing.

\section{Discussion}
 \label{sec:15N-Discussion}

   Observations indicate that \N14 \ production involves both,
   a primary and a secondary component (\eg, Matteucci 1986;
   Vila-Costas \& Edmunds 1993; van Zee \etal\ 1998),
   which is currently understood in terms of primary hydrostatic
   \N14 \ production in massive low metallicity stars
   (cf.\ Laird 1985, Timmes \etal\ 1995),
   while the secondary production of \N14 \ through the CNO cycle is
   dominating for $Z$ $\ga$ Z$_{\odot} / 100$
   (cf.\ Timmes \etal\ 1995; van den Hoek \& Groenewegen 1997).

   $^{15}$N is destroyed during hydrostatic hydrogen-burning
   (\eg, Wannier \etal\ 1991; El Eid 1994; Langer \& Henkel 1995) and
   is thus thought to be a product of explosive hydrogen nucleosynthesis.
   It is synthesized in novae as a mostly primary nucleus
   (\eg, Jos\'e \& Hernanz 1998).
   However, current models produce insufficient amounts to account
   for the galactic \N15 \ abundance (Kovetz \& Prialnik 1997).
   Alternatively, it may be produced in Type~II supernovae through
   neutrino spallation of primary \O16 \ (Woosley \& Weaver 1995,
   Timmes \etal\ 1995; see also Audouze \etal\ 1977).
   A small contribution from Type~Ia supernovae is also possible
   (Clayton \etal\ 1997; although see Nomoto \etal\ 1984).
   In the first case (novae), there is a time delay between star formation
   and the occurrence of \N15 \ enriched ejecta, leading to initially
   high \ISON\ ratios that gradually decrease with time
   (for an illustration, see \eg\ Fig.\,4 of G\"usten \& Ungerechts 1985).
   In the second case (massive stars), rapid injection of \N15 \ into the ISM
   leads to initially small \ISON\ ratios that gradually increase with time
   when the \N14 \ contribution from low-mass stars becomes significant.

   Analyzing \ISON\ data from HCN and \AMM, it soon became obvious that
   local ISM and galactic center \ISON\ ratios are moderately and
   strongly enhanced relative to the solar system value.
   Assuming that the solar system, the local ISM, and the galactic center
   region form a sequence of increasing degree of nuclear processing
   (the solar system was assumed to reflect the composition of the
   outer Galaxy at a time almost five billion years ago) led to the conclusion
   that \N15 \ is predominantly arising from massive stellar progenitors
   (\eg, Audouze \etal\ 1977; G\"usten \& Ungerechts 1985).
   In this case the more secondary \N14 \ nucleus becomes more abundant
   relative to \N15 \ with higher degree of nuclear processing,
   yielding \ISON\ ratios of 270, 300 -- 400, and 500 -- 1000 in the
   solar system, the local ISM, and the galactic center region, respectively.

   Two complications led to the present `nitrogen puzzle':
   Firstly, the Sun is more metal rich than the local ISM
   (\eg, Russell \& Dopita 1992; Cameron 1993; Cunha \& Lambert 1994;
   Meyer 1997), in spite of its age.
   Secondly, Dahmen \etal\ (1995) found a galactic disk gradient with
   \ISON\ ratios (from HCN) increasing with galactocentric distance.
   The first complication is easily resolved within the framework of
   primary \N15 \ production, if we note that the Sun is particularly
   enriched with nuclei processed in massive stars.
   This leads to enhanced \ISOB\ ratios (Henkel \& Mauersberger 1993;
   Henkel \etal\ 1994b) and to reduced \ISON\ ratios relative to the local ISM.
   The \ISON\ gradient is a more severe problem.
   The gradient implies that either the initial mass function (IMF) is biased
   in favor of high mass stars in the inner galactic disk
   (there is no observational support for such a bias) or that infall of halo
   gas or of cannibalized dwarf galaxies have altered interstellar abundances.
   Most galactic \ISON\ values are determined from
   \ISON\ : \ISOC\ double isotope ratios.
   Carbon fractionation may lead to small deviations between HCN/\HthCN\ and
   \ISOC\ abundance ratios (see Langer \etal\ 1984, 1989;
   Wilson \& Matteucci 1992; Wilson \& Rood 1994),
   but this should not alter the sign of a gradient.
   Assuming instead delayed injection of \N15 \ via novae into the ISM
   (\eg, Dahmen \etal\ 1995) fails to account for the large measured
   \ISON\ ratios in the galactic center region with its
   high number of potential nova candidates (see Shafter 1997).

   Solving the nitrogen puzzle requires the determination of small
   ($\approx$ 100) \ISON\ ratios in a (post-) starburst environment,
   that is strongly influenced by the ejecta from massive stars
   (Henkel \& Mauersberger 1993; Henkel \etal\ 1994b).
   Only a lower \ISON\ limit of order 100 was obtained toward M\,82
   (Henkel \etal\ 1998).
   Our \ISON\ ratio from NGC\,4945 strongly suggests
   that the bulk of \N15 \ is ejected by massive stars.

   The ISM of the LMC is less processed than the solar system
   and the local ISM and should also be characterized by
   overabundances of nuclei ejected from massive stars.
   [C/O]$_{\rm LMC}$ $\sim$ [N/O]$_{\rm LMC}$ $\sim -0.3$ is consistent with
   this hypothesis (\eg, Westerlund 1990).
   Thus, if \N15 \ is released by massive stars, an `overabundance' of
   \N15 \ relative to \N14 \ is expected;
   in the LMC, \ISON\ should be significantly smaller than
   ratios measured in the solar system and the local ISM.

   {\em Our observational results from the LMC and NGC\,4945 are
   thus providing strong support for \N15 \ being predominantly
   ejected by Type~II supernovae}.
   This finding is not consistent with the sign of
   the galactic disk \ISON\ gradient (Dahmen \etal\ 1995).
   It also seems to be in conflict with numerical calculations of massive star
   nucleosynthesis (\eg, Weaver \& Woosley 1993; Woosley \& Weaver 1995).
   While without the contribution of neutrino-induced nucleosynthesis to
   the production of \N15 , massive stars would destroy rather than produce
   this isotope (cf.\ Weaver \& Woosley 1993), Woosley \& Weaver (1995)
   find the neutrino production with the (uncertain) currently predicted
   neutrino scattering cross sections not sufficient to explain the
   solar system abundance of \N15 \ (cf.\ also Timmes \etal\ 1995).
   However, recent massive star models which take the effects of rotation
   on the stellar structure and nucleosynthesis into account
   (Heger \etal\ 1997, 1999, Langer \etal\ 1999) indicate the possibility
   of abundant hydrostatic production of \N15 \ in case of mixing
   between the hydrogen-burning and the helium-burning shell,
   a mechanism already discussed by Jorissen \& Arnould (1989).
   Consequently, although there are considerable uncertainties and
   weakly restricted parameters in the rotating massive star models
   which need to be explored in the near future, massive stars should be
   considered as an important source of \N15 \ in galaxies, particularly
   so after the observational facts reported in the present paper.

\acknowledgements

   We thank J.\,H.~Black and U.~Ott for useful discussions and criticism.
   YNC thanks for financial support through National Science Council
   of Taiwan grant \hbox{86-2112-M001-032}.
   RM acknowledges support from NATO grant SA.5-2-05 (CRG. 960086) 318/96,
   and NL from the Deutsche Forschungsgemeinschaft through
   grants La~587/15-1 and~16-1.

\vfil\eject

\figcaption[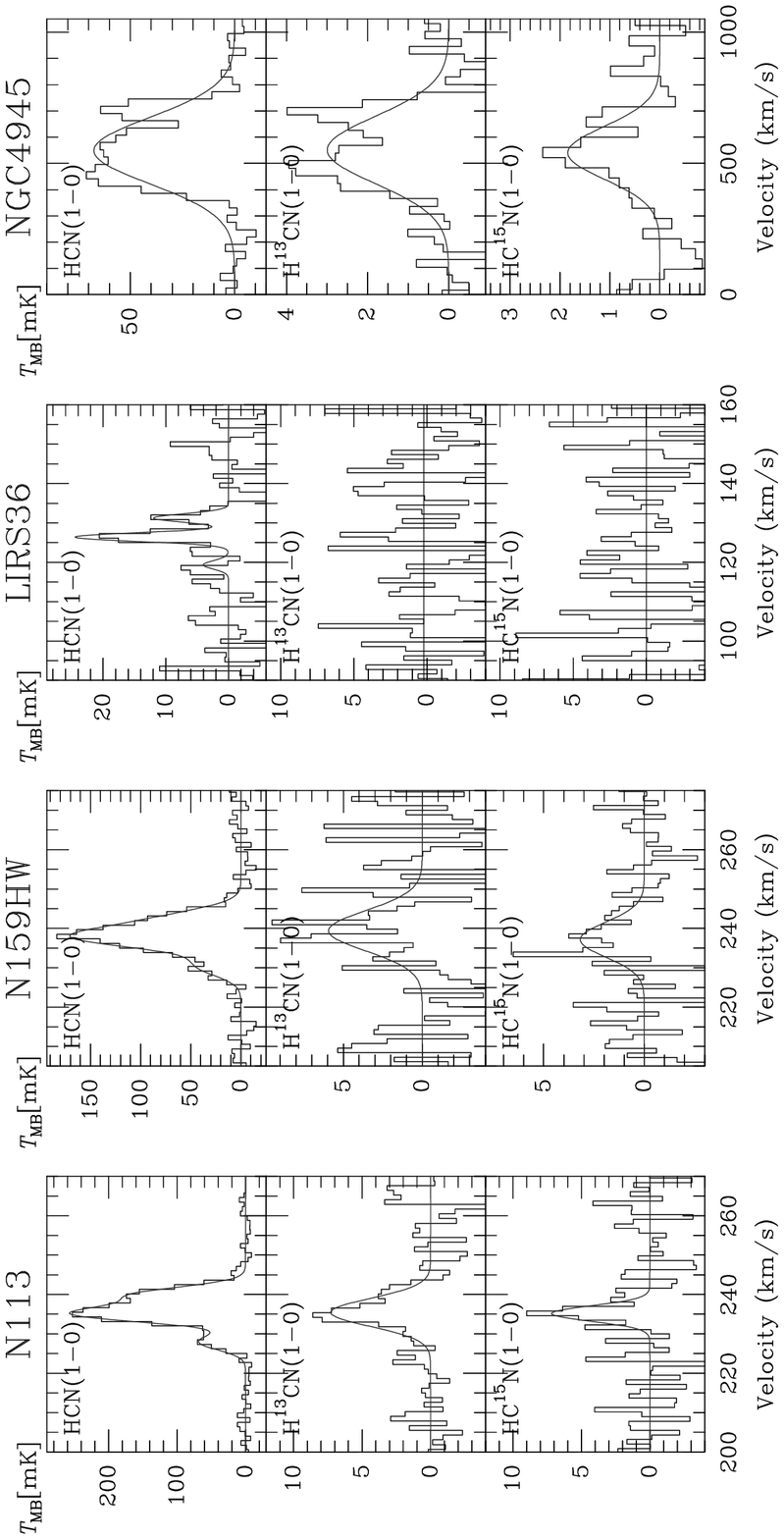]
           {Spectra of HCN, \HthCN, and \HCfiN; for coordinates, see
            Chin \etal\ (1997) and Henkel \etal\ (1994a)
 \label{fig:15N-spectra}}

\clearpage

\scriptsize
\begin{deluxetable}{l l r@{.}l r@{}l r c r@{.}l r@{.}l@{~$\pm$~}l}
\tablecolumns{9}
\tablecaption{Molecular lines parameters}
\scriptsize
\tablehead{
   \multicolumn{2}{l}{Molecule}
         & \multicolumn{2}{c}{\hspace{-3.2mm} Frequency}
         & \multicolumn{2}{l}{\TMB}   & r.m.s.~\tablenotemark{a}
         & \vLSR  \uspace1            & \multicolumn{2}{c}{\delv}
         & \multicolumn{3}{c}{$\int$ \TMB\,d$v$} \\
   \multicolumn{2}{l}{\& Transition}
         & \multicolumn{2}{c}{\hspace{-3.2mm} [GHz]}
         & \multicolumn{2}{l}{[mK]}   & [mK]
         & [\kms] \dspace             & \multicolumn{2}{c}{[\kms]}
         & \multicolumn{3}{c}{[\Kkms]}           \\
}
\startdata
   {\bf N113} \\
   HCN \tablenotemark{b}
                  & $J$=1--0 \bra2 \begin{tabular}{l}
                    $F$=1--1  \\ $F$=2--1  \\ $F$=0--1  \end{tabular}
                  & \multicolumn{2}{c}{\hspace{-0.9eM} \begin{tabular}{r@{.}l}
                                    88&630416   \hspace{-1eM}\\
                                    88&631847   \\   88&633936 \end{tabular}}
                  & \begin{tabular}{r}  158  \\  246  \\   68  \end{tabular}
                    \hspace*{-3.4mm} &
                           &  11 &  235.0 &  4&8 & \multicolumn{3}{c}{
                    \begin{tabular}{r@{.}l@{~$\pm$~}l}  0&809 & 0.057 \\
                      1&26  & 0.06  \\ 0&346 & 0.057 \end{tabular}}\\
   \HthCN\ \tablenotemark{c}
                  & $J$=1--0     &  86&340184
                  &   7&.3 &   3 &  235.4 &  8&3 & 0&065 & 0.004 \\
   \HCfiN\ \tablenotemark{c}
                  & $J$=1--0     &  86&054961
                  &   7&.2 &   3 &  235.2 &  4&6 & 0&035 & 0.004 \\
\uspace2
   {\bf N159HW} \\
   HCN \tablenotemark{b}
                  & $J$=1--0 \bra2 \begin{tabular}{l}
                    $F$=1--1  \\ $F$=2--1  \\ $F$=0--1  \end{tabular}
                  & \multicolumn{2}{c}{\hspace{-0.9eM} \begin{tabular}{r@{.}l}
                                    88&630416   \hspace{-1eM}\\
                                    88&631847   \\   88&633936 \end{tabular}}
                  & \begin{tabular}{r}   65  \\  156  \\   43  \end{tabular}
                    \hspace*{-3.4mm} &
                           &  13 &  237.7 &  6&2 & \multicolumn{3}{c}{
                    \begin{tabular}{r@{.}l@{~$\pm$~}l}  0&425 & 0.087 \\
                      1&03  & 0.09  \\ 0&283 & 0.087 \end{tabular}}\\
   \HthCN\ \tablenotemark{c}
                  & $J$=1--0     &  86&340184
                  &   5&.9 &   5 &  239.4 & 12&0 & 0&076 & 0.009 \\
   \HCfiN\ \tablenotemark{c}
                  & $J$=1--0     &  86&054961
                  &   3&.2 &   3 &  237.2 & 10&4 & 0&035 & 0.005 \\
\uspace2
   {\bf LIRS\,36} \\
   HCN \tablenotemark{b}
                  & $J$=1--0 \bra2 \begin{tabular}{l}
                    $F$=1--1  \\ $F$=2--1  \\ $F$=0--1  \end{tabular}
                  & \multicolumn{2}{c}{\hspace{-0.9eM} \begin{tabular}{r@{.}l}
                                    88&630416   \hspace{-1eM}\\
                                    88&631847   \\   88&633936 \end{tabular}}
                  & \begin{tabular}{r}   12  \\   25  \\    4  \end{tabular}
                    \hspace*{-3.4mm} &
                           &   7 &  126.4 &  2&5 & \multicolumn{3}{c}{
                    \begin{tabular}{r@{.}l@{~$\pm$~}l}  0&032 & 0.018 \\
                      0&066 & 0.019 \\ 0&011 & 0.018 \end{tabular}}\\
   \HthCN\ \tablenotemark{c}
                  & $J$=1--0     &  86&340184
                  & $<$ 5& &   6 &   ---  & \multicolumn{2}{l}{~~---}
                         & \multicolumn{3}{c}{$<$ 0.020 \tablenotemark{d}} \\
   \HCfiN\ \tablenotemark{c}
                  & $J$=1--0     &  86&054961
                  & $<$ 5& &   6 &   ---  & \multicolumn{2}{l}{~~---}
                         & \multicolumn{3}{c}{$<$ 0.024 \tablenotemark{d}} \\
\uspace2
   {\bf NGC\,4945} \\
   HCN            & $J$=1--0     &  88&631847
                  &  67&   &  11 &  550   & \multicolumn{2}{l}{~~300}
                                                & 21&3   & 0.1   \\
   \HthCN\        & $J$=1--0     &  86&340184
                  &   3&.0 &   1 &  550   & \multicolumn{2}{l}{~~300}
                                                &  0&959 & 0.013 \\
   \HCfiN\        & $J$=1--0     &  86&054961
                  &   1&.8 &   1 &  540   & \multicolumn{2}{l}{~~250}
                                                &  0&464 & 0.012 \\
\enddata
 \label{tbl:15N-parameter}
\tablenotetext{a}
    {1\,$\sigma$ noise level of a single channel width of 0.14\,\kms\ (for the
     Magellanic Clouds) or 2.4\,\kms\ (for NGC\,4945) on a \TMB\ scale.}
\tablenotetext{b}
    {The three HCN hyperfine transitions ($F$=1--1, $F$=2--1, $F$=0--1)
     have been resolved by a Gaussian fit.
     While \TMB\ values and integrated line intensities are given for
     each component, the velocity refers to the main component.}
\tablenotetext{c}
    {The hyperfine components of \HthCN\ and \HCfiN\ cannot be resolved.}
\tablenotetext{d}
    {For undetected lines, 3\,$\sigma$ values are given.}
\end{deluxetable}

\clearpage

\scriptsize
\begin{deluxetable}{l r@{~$\pm$~}l r@{~$\pm$~}l r@{~$\pm$~}l r@{~$\pm$~}l
                     r@{~$\pm$~}l c r@{~$\pm$~}l r@{~$\pm$~}l}
\tablecolumns{9}
\tablewidth{0pc}
\tablecaption{Integrated line intensity ratios `$R$' of HCN isotopic species,
              integrated line intensity ratios `$R_{02}$' and `$R_{12}$'
              of HCN hyperfine transitions with opacity $\tau_0$, and isotopic
              abundance ratios.}
\tablehead{
   Source & \multicolumn{2}{c}{$R \left( \frac{\hbox{HCN}}{\HthCN} \right)$}
          & \multicolumn{2}{c}{$R \left( \frac{\hbox{HCN}}{\HCfiN} \right)$}
          & \multicolumn{2}{c}{$R \left( \frac{\HthCN}{\HCfiN} \right)$}
          & \multicolumn{2}{c}{$R_{12}$\tablenotemark{a}}
          & \multicolumn{2}{c}{$R_{02}$\tablenotemark{a}}
          & $\tau_0$\tablenotemark{a}
          & \multicolumn{2}{c}{\ISOC} & \multicolumn{2}{c}{\ISON} \\
}
\startdata
   N113      &~~~37  &   5  &~~~69  &  13  &~~~1.9  & 0.3
             &  0.64 & 0.07 &  0.27 & 0.06 & 0.122  & ~~~60 & 6 & ~111 & 17 \\
   N159HW    &   23  &   6  &   49  &  15  &   2.1  & 0.6
             &  0.41 & 0.12 &  0.27 & 0.11 & $< 1$
                       & \multicolumn{2}{c}{$\approx 50$ \tablenotemark{b}}
                       & \multicolumn{2}{c}{$\approx 100$ \tablenotemark{c}} \\
   LIRS\,36  & \multicolumn{2}{c}{$>$ 4.5} & \multicolumn{2}{c}{$>$ 5.5}
             & \multicolumn{2}{c}{---}
             &  0.49 & 0.41 &  0.17 & 0.32 &  ---
             & \multicolumn{2}{c}{---} & \multicolumn{2}{c}{---} \\
   NGC\,4945 &   22  &   1  &   46  &   1  &   2.1  & 0.1
             & \multicolumn{2}{c}{---} & \multicolumn{2}{c}{---}
             & $< 1$   & \multicolumn{2}{c}{$\approx 50$ \tablenotemark{b}}
                       & \multicolumn{2}{c}{$\approx 100$ \tablenotemark{c}} \\
\enddata
 \label{tbl:15N-ratio}
\tablenotetext{a}
    {The opacity of the HCN $J$=1--0 transition can be derived from the
     line intensity ratios of the $J$=1--0 hyperfine components.
     The degeneracies of the HCN $J=1$, $F=0, 1, 2$ states are 1, 3, and 5.
     If $\tau_0$, $\tau_1$, $\tau_2$ denote the optical depths of the
     transitions from these three states to the ground state, respectively,
     we find under conditions of Local Thermodynamical Equilibrium (LTE)
     $\tau_1$ = 3\,$\tau_0$ and $\tau_2$ = 5\,$\tau_0$.
     The intensity ratios of the hyperfine transitions are then
     $R_{12} = \frac{\hbox{$I$($F$=1--1)}}{\hbox{$I$($F$=2--1)}}
     = \frac{\hbox{$1-{\rm e}^{-3\tau_0}$}}{\hbox{$1-{\rm e}^{-5\tau_0}$}}$ and
     $R_{02} = \frac{\hbox{$I$($F$=0--1)}}{\hbox{$I$($F$=2--1)}}
     = \frac{\hbox{$1-{\rm e}^{-\tau_0}$}}{\hbox{$1-{\rm e}^{-5\tau_0}$}}$.
     For NGC\,4945, $\tau_0$ was estimated from \ISOC\ (Henkel \etal\ 1994a) and
     the observed line intensity ratio (see Table\,\ref{tbl:15N-parameter}).
    }
\tablenotetext{b}
    {For the \ISOC\ ratio in N159, see Johansson \etal\ (1984);
     for NGC\,4945, see Henkel \etal\ (1994a).
    }
\tablenotetext{c}
    {For NGC\,4945, lines are too broad to see individual hyperfine
     components in the HCN line, but the \ISON\ ratio can be estimated by
     \ISON\ = $\frac{\HthCN}{\HCfiN}$ $\times$ \ISOC.
     This can also be used to confirm the \ISON\ ratio determined in N159HW.
     }
\end{deluxetable}

\plotone{15N.spectra.ps}


\begin{thebibliography}{}
 \bibitem[]{}
   Audouze, J., Lequeux, J., Rocca-Volmerange, \& B., Vigroux, L. 1977,
   CNO Isotopes in Astrophysics, ed. J. Audouze, D. Reidel Publishing Company,
   Dordrecht, p155
 \bibitem[1997]{}
  Cameron, A.G.W. 1993, Protostars \& Planets III, eds. E.H. Levy, J.I. Lunine,
  The University of Arizona Press, Tucson, p47
 \bibitem[]{}
   Chin, Y.-N., Henkel, C., Millar, T.J., Whiteoak, J.B., \& Mauersberger, R.,
   1996, A\&A, 312, L33
 \bibitem[1997]{}
   Chin, Y.-N., Henkel, C., Whiteoak, J.B., \etal\ 1997, A\&A, 317, 548
 \bibitem[1997]{}
   Chin, Y.-N., Henkel, C., Millar, T.J., Whiteoak, J.B., \& Marx-Zimmer, M.
     1998, A\&A, 330, 901
 \bibitem[1997]{}
   Clayton, D.D., Arnett, D., Kane, J., \& Meyer, B.S. 1997, ApJ, 486, 824
 \bibitem[1997]{}
   Cunha, K., \& Lambert, D.L. 1994, ApJ, 426, 170
 \bibitem[1997]{}
   Dahmen, G., Wilson, T.L., \& Matteucci, F. 1995, A\&A, 295, 194
 \bibitem[]{}
   El Eid, M.F. 1994, A\&A, 285, 915
 \bibitem[]{}
   G\"usten, R., \& Ungerechts, H. 1985, A\&A, 145, 241
 \bibitem[]{}
   Heger, A., Jeannin, L., Langer, N., \& Baraffe, I. 1997, A\&A, 327, 224
 \bibitem[]{}
   Heger, A., Woosley, S.E., \& Langer, N. 1999, ApJ, in preparation
 \bibitem[]{}
   Heikkil\"a, A., Johansson, L.E.B., \& Olofsson, H. 1997, A\&A, 319, L21
 \bibitem[]{}
   Heikkil\"a, A., Johansson, L.E.B., \& Olofsson, H. 1998, A\&A, 332, 493
 \bibitem[]{}
   Henkel, C., \& Mauersberger, R. 1993, A\&A, 274, 730
 \bibitem[]{}
   Henkel, C., Whiteoak, J.B., \& Mauersberger, R. 1994a, A\&A, 284, 17
 \bibitem[]{}
   Henkel, C., Wilson, T.L., Langer, N., Chin, Y.-N., \& Mauersberger, R. 1994b,
     The Structure and Content of Molecular Clouds, eds. T.L. Wilson, K.J.
     Johnston, Springer Verlag, Berlin, p72
 \bibitem[]{}
   Henkel, C., Chin, Y.-N., Mauersberger, R., \& Whiteoak, J.B. 1998,
     A\&A, 329, 443
 \bibitem[]{}
   Johansson, L.E.B., Olofsson, H., Hjalmarson, \AA, Gredel, R., \& Black, J.H.,
     1994, A\&A, 291, 89
 \bibitem[]{}
   Jorissen, A., \& Arnould, M. 1989, A\&A, 221, 161
 \bibitem[]{}
   Jos\'e, J., \& Hernanz, M. 1998, ApJ, 494, 680
 \bibitem[]{}
   Koornneef, J. 1993, ApJ, 403, 581
 \bibitem[]{}
   Kovetz, A., \& Prialnik, D. 1997, ApJ, 477, 356
 \bibitem[]{}
   Laird, J.B. 1985, ApJ, 289, 556
 \bibitem[]{}
   Langer, N., \& Henkel, C. 1995, Sp. Sci. Rev., 74, 343
 \bibitem[]{}
   Langer, N., Heger, A., \& Garc\'{\i}a-Segura, G. 1999, Rev. Modern Astron.,
     Vol.~11, in press
 \bibitem[]{}
   Langer, W.D., \& Graedel, T.E. 1989, ApJS, 69, 241
 \bibitem[]{}
   Langer, W.D., Graedel, T.E., Frerking, M.A., \& Armentrout, P.B. 1984,
     ApJ, 277, 581
 \bibitem[]{}
   Lovas, F.J. 1992, J. Phys. Chem. Ref. Data, 21, 181
 \bibitem[]{}
   Matteucci, F. 1986, MNRAS, 221, 911
 \bibitem[]{}
   Meyer, D.M. 1997, IAU Symp. 178, Molecules in Astrophysics: Probes and
      Processes, ed. E.F. van Dishoeck, Kluwer Academic Publishers, Dordrecht,
      p407
 \bibitem[]{}
   Nomoto, K., Thielemann, F.-K., \& Yokoi, K. 1984, ApJ, 286, 644
 \bibitem[]{}
   Prantzos, N., Aubert, O., \& Audouze, J. 1996, A\&A, 309, 760
 \bibitem[]{}
   Russell, S.C., \& Dopita, M.A. 1992, ApJ, 384, 508
 \bibitem[]{}
   Shafter, A.W. 1997, ApJ, 487, 226
 \bibitem[]{}
   Timmes, F.X., Woosley, S.E., \& Weaver, T.A. 1995, ApJS, 98, 617
 \bibitem[]{}
   van den Hoek, L.B., \& Groenewegen, M.A.T. 1997, A\&AS, 123, 305
 \bibitem[]{}
   van Zee, L., Salzer, J.J., \& Haynes, M.P. 1998, ApJ, 497, L1
 \bibitem[]{}
   Vila-Costas, M.B., \& Edmunds, M.G. 1993, MNRAS, 265, 199
 \bibitem[]{}
   Wannier, P.G., Linke, R.A., \& Penzias, A.A. 1981, ApJ, 247, 522
 \bibitem[]{}
   Wannier, P.G., Andersson, B.-G., Olofsson, H., Ukita, N., \& Young, K. 1991,
     ApJ, 380, 593
 \bibitem[]{}
   Watson, W.D., Anicich, V.G., \& Huntress, W.T., Jr. 1976, ApJ, 205, L165
 \bibitem[]{}
   Weaver, T.A., \& Woosley, S.E. 1993, Phys. Rep., 227, 65
 \bibitem[]{}
   Westerlund, B.E. 1990, A\&AR, 2, 29
 \bibitem[]{}
   Wilson, T.L., \& Matteucci, F. 1992, A\&AR, 4, 1
 \bibitem[]{}
  Wilson, T.L., \& Rood, R.T. 1994, ARAA, 32, 191
 \bibitem[]{}
   Woosley, S.E., \& Weaver, T.A. 1995, ApJS, 101, 181
\end{thebibliography}
\end{document}